\documentclass[twocolumn,amssymb,amsmath,floats,showpacs,pre]{revtex4-1}

\usepackage{amsmath,amssymb,bm}
\usepackage{graphicx}
\usepackage{xcolor}

\begin{document}

\title{Cooperative epidemics on multiplex networks
}

\author{N. Azimi-Tafreshi}

\affiliation{Physics Department, Institute for Advanced Studies in Basic Sciences, 45195-1159 Zanjan, Iran }


\begin{abstract}
The spread of one disease, in some cases, can stimulate the spreading of another infectious disease. Here, we treat analytically
a symmetric co-infection model for spreading of two diseases on a two-layer multiplex network. We allow layer overlapping, but we
assume that each layer is random and locally loop-less. Infection with one of the diseases increases the probability of getting infected with the other. Using the generating function method, we calculate exactly the fraction of individuals
infected with both diseases (so-called co-infected clusters) in the stationary state, as well as the epidemic spreading
thresholds and the phase diagram of the model. With increasing cooperation, we observe a tricritical point and the type of
transition changes from continuous to hybrid. Finally, we compare the co-infected clusters in the case of co-operating diseases
with the so-called ``viable" clusters in networks with dependencies.
\end{abstract}


\maketitle

\section{Introduction}
Cooperation between two epidemics occurs when the spread of one disease increases the spreading of the other. It was estimated in 2011 \cite{TB} that, about one-third of a total world population of 7 billion people had active
tuberculosis (TB). But if one restricted oneself to the $\sim 33$ million people with human immunodeficiency virus (HIV),
about 30\% also had active or latent TB.
Inversely, of the active TB cases, about 15 \% also had HIV, which is nearly a factor of 100 higher than the incidence rate in
the total population. People with HIV and TB co-infection typically also experience faster disease progression than those with
TB or HIV alone. Another dramatic example where two diseases mutually enhance their spreading is HIV and hepatitis
C virus (HCV)\cite{HIV-HCV}. Also there, it is estimated that about one third of all people with HIV are also co-infected with
HCV.

These numbers show already that mutual co-infection is a huge problem. But there are also other recent examples like HIV and
hepatitis B \cite{Rockstroh}, while historically the case of Spanish flu and TB was one of the most devastating \cite{Brundage}.

Therefore, much effort recently has been devoted to studying the dynamics of spreading of two or more co-operating pathogens
\cite{immune,Interaction,Sanz,Fakhteh,Cai,clustercoin,Hanna,timescale}. Let us discuss just
a few of these papers. Newman {\it et al.} \cite{Interaction} assumed an asymmetric
rule for cooperation, such that the first disease spreads independently of the second one, but the second can propagate only
among those that had already been infected with the first disease. This simplifies the analysis, of course, but prevents the
most dramatic scenarios that may occur if the cooperation is mutual and symmetric, as assumed in
\cite{Sanz,Fakhteh,Cai}. Sanz {\it et al.} \cite{Sanz} proposed a framework, based on the heterogeneous mean-field approximation, for the spontaneous spreading of two diseases and studied different effects of one disease on the spreading of the other. Also, in \cite{Fakhteh} and \cite{Cai}, each disease can spread independently of the other, but
the secondary infection rates are enhanced compared to the rates for infection by the first diseases. Such a model was treated in mean field theory in \cite{Fakhteh}, while detailed simulations in various types of networks are reported in
\cite{Cai}.

The main result in \cite{Cai} was that the typical ``continuous" phase transition observed in simple epidemic models can be
replaced -- depending on details of the networks and of the infection processes -- by ``discontinuous" ones, where the
incidence rate at threshold does not increase continuously but jumps immediately
to a finite value. In typical discontinuous (or ``first order") phase transitions, there is no sign of warning -- like enhanced
fluctuations, an increasing correlation length or a slowing down of the dynamics, which occur in single epidemics \cite{SIRbond3} --
as the threshold for large-scale spreading
is approached. This is of course the most worrying aspect for health policies, but fortunately most of the discontinuous
transitions found in \cite{Cai} were ``hybrid", i.e. they combined the jumps of first order transitions with the
anomalies in the approach to the threshold seen in continuous (or ``second order") transitions.

The results in  \cite{immune,Interaction,Sanz,Fakhteh,Cai,clustercoin,Hanna,timescale}
are extremely interesting, but most were obtained either by some sort of mean
field theory (i.e., all network properties were neglected and/or stochastic fluctuations were assumed to be absent) or
by simulating very specific cases.
The range of phenomena found in \cite{Cai} strongly suggests that one should look for analytic results that do take
into account fluctuations and at least some simple network structure.

This is the main aim of the present paper. Another aim is to understand the links between co-infections and
catastrophic cascades in networks with interdependencies \cite{Buldyrev}. The latter can be understood
most easily \cite{Son, avalanche, Min, multidir, Grassberger}
as ``viable" clusters in multiplex networks. Multiplex networks \cite{multilayer,multiplex, multiplex2} have nodes of one type and multiple types of edges. They can be treated as a superposition of several network layers, where nodes are coupled to their counterparts in different layers. Hence multiplex networks can be represented as edge-colored multigraphs, where each edge color corresponds to a network layer. Here we consider a multiplex network with two types of edges, such that each type of edge allows for spreading of one of two types of agents. A cluster in such a network is called viable \cite{avalanche}, if each of its nodes can be reached from any other node by both types of agents.
Obviously there is an analogy with cooperating coinfections, if we identify the two agents with the two
pathogens: In the limit of strong cooperativity, large infected clusters will always be coinfected, i.e. each node
on such a cluster will be reachable by both pathogens propagating only on the cluster. But the
detailed nature of this connection has remained elusive up to now. It is clarified in the present paper.

Finally, we should also point out that the case of co-operating epidemics is very much different from the case
of competing or antagonistic epidemics \cite{newman, Funk, Funk1, competition, idea,modeling,awareness,sahneh,asymetry,memes}.
Although the latter are also of huge practical interest, their
dynamics is very different and leads in general to less dramatic effects.

The paper is organized as follows. In Sec. II, we introduce a co-infection model on a multiplex network with two types of edges and present an analytical framework enabling us to describe the nature of the transitions corresponding to the emergence of co-infected clusters. We apply our general results to the Erd\H{o}s--R\'enyi multiplex networks. The paper is concluded in Sec. III.

\section{Cooperative epidemics}

\begin{figure*}
\begin{center}
\scalebox{0.5}{\includegraphics[angle=0]{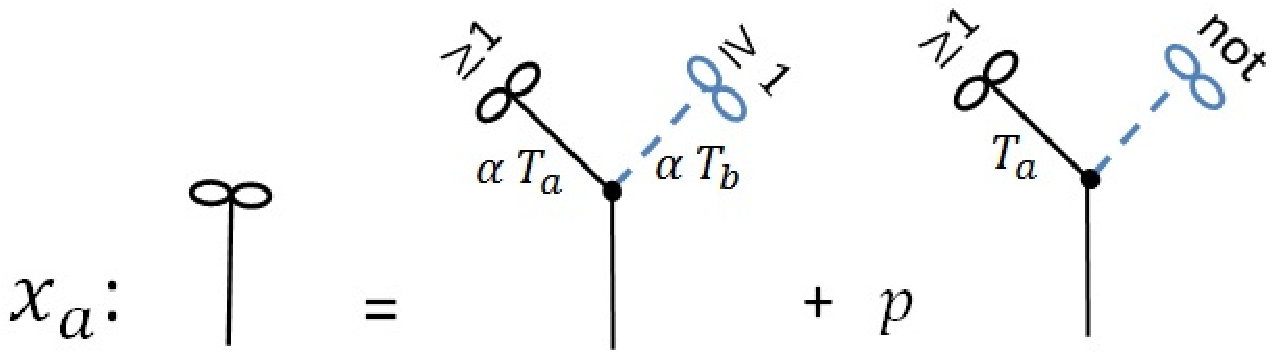}}~~~~
\scalebox{0.5}{\includegraphics[angle=0]{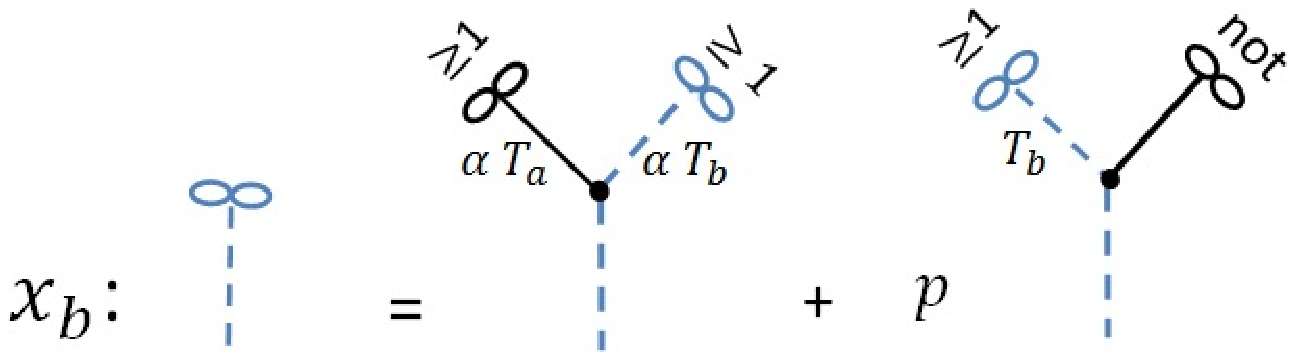}}
\end{center}
\caption{(color online). Schematic representation of the self-consistency equations for the probabilities $x_a$ and $x_b$. Solid
black and dashed blue lines with infinity symbols at one end
represent probabilities $x_a$ and $x_b$, respectively. For the sake of clarity, we do
not show the edges leading to finite components, namely probabilities $1-x_a$ and $1-x_b$.
}
\label{f1}
\end{figure*}
\begin{figure}[t]
\begin{center}
\scalebox{0.55}{\includegraphics[angle=0]{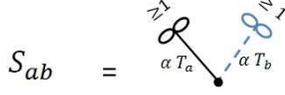}}
\end{center}
\caption{(color online). Schematic representation of the probability that a node belongs to the giant co-infected cluster.
}
\label{f2}
\end{figure}
\subsection{Analytical framework}

Let us consider an uncorrelated multiplex network having two types of edges $i=a,b$. The network can be treated as a superposition of two network layers with edges of type $a$ and $b$, such that overlapping of two types of edges can exist for some pairs of nodes. We define degree $k_i$, as the number of edges of type $i$ that are incident to a node and $k_{ab}$ denotes the number of overlapped edges of each node. The multiplex network is completely described by the joint degree distribution $P(k_a, k_b, k_{ab})$, and we assume that each layer has a locally treelike structure in the infinite network limit.

A co-infection model is defined for two diseases $a$ and $b$, spreading on the multiplex network. Each of $i$ diseases spreads with transmission probability $T_i$ through only edges of type $i$, while the overlapped edges can transmit both diseases with probability $T_{ab}$. We assume that both diseases follow the susceptible$-$infective$-$recovered (SIR) dynamics \cite{kermack}. A given random node can be infected with disease $i$, if it has at least one edge of type $i$, connecting it to its infected neighbors. We assume that during the spreading process, if a node can receive both diseases, each one through at least one edge of each type $i$, it receives the diseases with a higher probability, such that the transmissibilities $T_{a}$ and $T_{b}$ are increased by factor $\alpha >1$. Also, through the overlapped edges both diseases can be transmitted with the enhanced transmissibility $T_{ab}> T_a T_b$.

It was shown that there is a mapping between the SIR epidemic model and the bond percolation theory, such that the set of individuals infected by a disease outbreak with transmissibility $T$, has the same size as the giant connected cluster of occupied edges with occupation probability $T$ \cite{SIRbond1, SIRbond3}. The mapping can be extended to multiplex networks and one can treat percolation on multiplex networks as an epidemic spreading process \cite{Son}. Each edge of type $i$ is occupied with probability $T_i$, equal to the probability that the end node of that edge (the neighbor of an infected node) will become infected with disease $i$. Here, we consider the cooperation between two diseases. If a given node has at least one edge of each type, connecting the node to the infinite infected clusters, the transmissibility of these edges is enhanced. The probability that a node belongs to the giant co-infected cluster is equal to the probability that a given node has enhanced occupied edges of both types, connecting it to the giant connected cluster of enhanced occupied edges. Hence, using percolation theory and the generation function method, we can solve exactly for the fraction of individuals, infected with both diseases in configuration model networks with arbitrary degree distributions.

Those nodes connected with the occupied overlapped edges can behave as a whole, since if one of them is infected by both diseases, all others will also be infected. Hence we can merge theses nodes into a single node, a so-called supernode \cite{Havlin, linkoverlap}. In other words, the network is renormalized to a network with supernodes connected with only the non-overlapped edges. One can find the probability that a random node belongs to a supernode of size $m$, denoted by $R(m, T_{ab})$ \cite{componentsize}. Assuming that there is no correlation between the overlapped and non-overlapped edges in the original network, namely, $P(k_a, k_b, k_{ab})=P(k_a,k_b)P(k_{ab})$, the size distribution of the supernodes is obtained for every arbitrary overlapping degree distribution $P(k_{ab})$,
\begin{eqnarray}
R(m, T_{ab})=\frac{T_{ab}^{m-1}\langle k_{ab}\rangle}{(m-1)!}\Big[\frac{d^{m-2}}{dx^{m-2}}[G_1(x)]^m\Big]_{x=1-T_{ab}}\nonumber\\
\label{eq101}
\!\!\!\!\!
\end{eqnarray}
in which $G_1(x)$ is the generation function for the distribution of the overlapped degrees of nodes, reached by following a randomly chosen overlapped edge and is given by \cite{GF}
\begin{eqnarray}
G_1(x)= \sum_{k_{ab}} \frac{k_{ab} P(k_{ab})}{\langle k_{ab} \rangle}x^{k_{ab}-1}.
\label{eq2}
\!\!\!\!\!
\end{eqnarray}

We define $\mathbf{q}\equiv (q_a, q_b)$ as the degree of supernodes which denotes the number of non-overlapped edges of each supernode. From renormalization theory, the degree distribution of supernodes of size $m$, $P_{m}(\mathbf{q})$ is determined as the distribution of the sum of $m$ random variables chosen from the marginal (non-overlapped) degree distribution $P(k_a,k_b)$, which is the $m$th-order convolution of $P(k_a,k_b)$ \cite{SRG}.

To find the size of the giant co-infected cluster, for each type $i=a,b$ of edge we define $x_i$ to be the probability that the end node (supernode) of a randomly chosen edge of type $i$ is the root of an infinite sub-tree infected with disease $i$. The subtree infected with disease $i$, by definition means that the subtree's nodes have disease $i$, but they can have the other disease or not. The probabilities $x_a$ and $x_b$, are schematically shown in Fig.~\ref{f1}. These probabilities play the
role of the order parameters of the phase transition associated
with the emergence of the giant co-infected cluster. We can write the self-consistency
equations for probabilities $x_i$ using the locally
treelike structure of the renormalized networks,
\begin{eqnarray}
x_i=R_{\infty}+&&\sum_{m=1}^{\infty}R(m, T_{ab})\sum_\textbf{q}\frac{q_i P_m(\textbf{q})}{\langle q_i\rangle}\times\nonumber\\
&& \Bigl{(}\Big[1-(1-\alpha T_i x_i)^{q_i-1}\Big]\Big[1-(1-\alpha T_j x_j)^{q_j}\Big]\nonumber\\
&&+p \Big[1-(1- T_i x_i)^{q_i-1}\Big](1- T_j x_j)^{q_j}\Bigr{)}.\nonumber\\
\label{x_i}
\!\!\!\!\!
\end{eqnarray}
Where $R_{\infty}$ is the probability that a given node belongs to a supernode of infinite size.

Let us explain the right-hand terms in Eq.~(\ref{x_i}). The probability that the end node (supernode) of a randomly chosen edge of type $i$, has degree $q_i$ is $q_{i}P_{m}(\textbf{q})/\langle q_{i}\rangle $. There are two possibilities:
If the end node of the randomly chosen edge of type $i$, has at least one edge of each type $i$, which leads to an infected cluster with disease $i$ (probability $x_i$), the transmissibility of each edge is increased by a factor $\alpha$. The second line of Eq.~(\ref{x_i}), indicates the contribution of this possibility. Setting $T_i=1$ and $\alpha=1$, this part gives us the size of the giant viable cluster \cite{Son, avalanche}.
The second possibility is when the end node is only connected to type $i$ edges, leading to an infinite infected subtree of type $i$. In this case, there is no cooperation between two diseases. The third line in Eq.~\ref{x_i} is related to this possibility. Note that these terms do not make any contribution toward deriving the giant viable clusters. In order to compare the giant coinfected and the viable clusters, we add the contribution of these terms by factor $p$, which is equal to 1 in our co-infection model and is 0 for the viable clusters.

Using these probabilities, we can obtain the probability that a randomly chosen node belongs to the giant co-infected cluster, denoted by $S_{ab}$. This probability, shown schematically in Fig~\ref{f2}, is given by the following expression,
\begin{eqnarray}
S_{ab}=R_{\infty}+&&\sum_{m=1}^{\infty}R(m, T_{ab})\sum_\textbf{q}P_m(\textbf{q})\times\nonumber\\
&&\Big[1-(1-\alpha T_a x_a)^{q_a}\Big]\Big[1-(1-\alpha T_b x_b)^{q_b}\Big].\nonumber\\
\label{eq4}
\!\!\!\!\!
\end{eqnarray}
We can rewrite Eqs.~(\ref{x_i}) and (\ref{eq4}) in terms of the generating functions of each network as,
\begin{eqnarray}
x_i=R_{\infty}+&&\sum_{m=1}^{\infty}R(m, T_{ab})\times\nonumber\\
&& \Bigl{(}\Big[1-G_{1}^{i}(1-\alpha T_i x_i)\Big]\Big[1-G_{0}^j(1-\alpha T_j x_j)\Big]\nonumber\\
&&+p \Big[1-G_{1}^{i}(1- T_i x_i)\Big]G_{0}^j(1- T_j x_j)\Bigr{)}.
\label{eq5}
\!\!\!\!\!
\end{eqnarray}
and
\begin{eqnarray}
S_{ab}=R_{\infty}+&&\sum_{m=1}^{\infty}R(m, T_{ab})\times\nonumber\\
&&\Big[1-G_{0}^a(1-\alpha T_a x_a)\Big]\Big[1-G_{0}^b(1-\alpha T_b x_b)\Big].\nonumber\\
\label{Sab}
\!\!\!\!\!
\end{eqnarray}
where $G_0^i(x)$ and $G_1^i(x)$ are the generating functions for the degree distribution and the excess degree distribution, respectively:
\begin{equation}
G_0^i(x)\equiv \sum_{q_i} P(q_i)x^{q_i},~~G_1^i(x)=\sum_{q_i}\frac{q_{i} P(q_i)}{\langle q_i \rangle}x^{q_i-1}
\label{g}
.
\end{equation}
Index $i$ for the generation functions refers to types of edges $i=a,b$.
\begin{figure*}[t]
\begin{center}
\scalebox{0.43}{\includegraphics[angle=0]{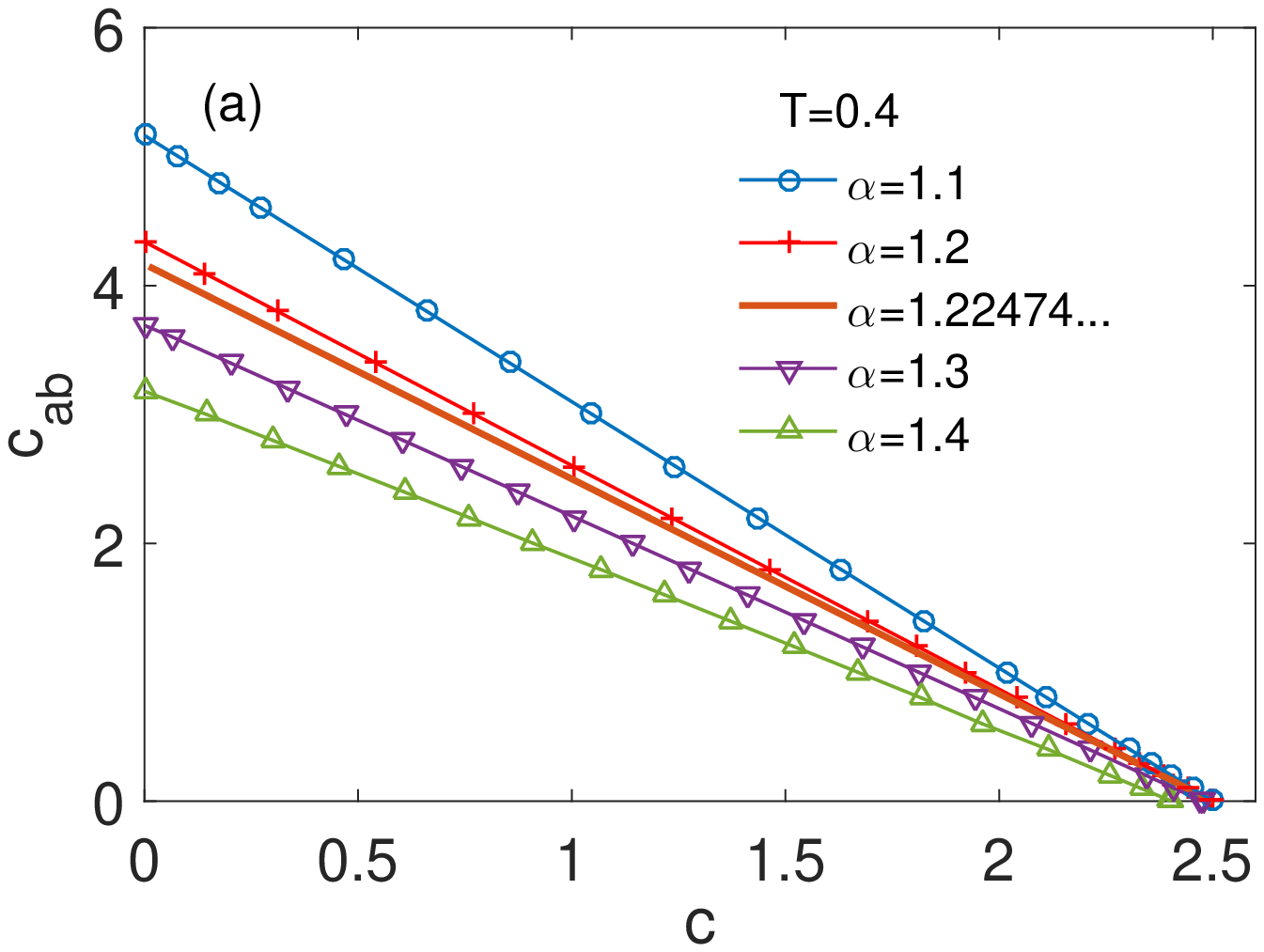}}~~~~~~~~~
\scalebox{0.43}{\includegraphics[angle=0]{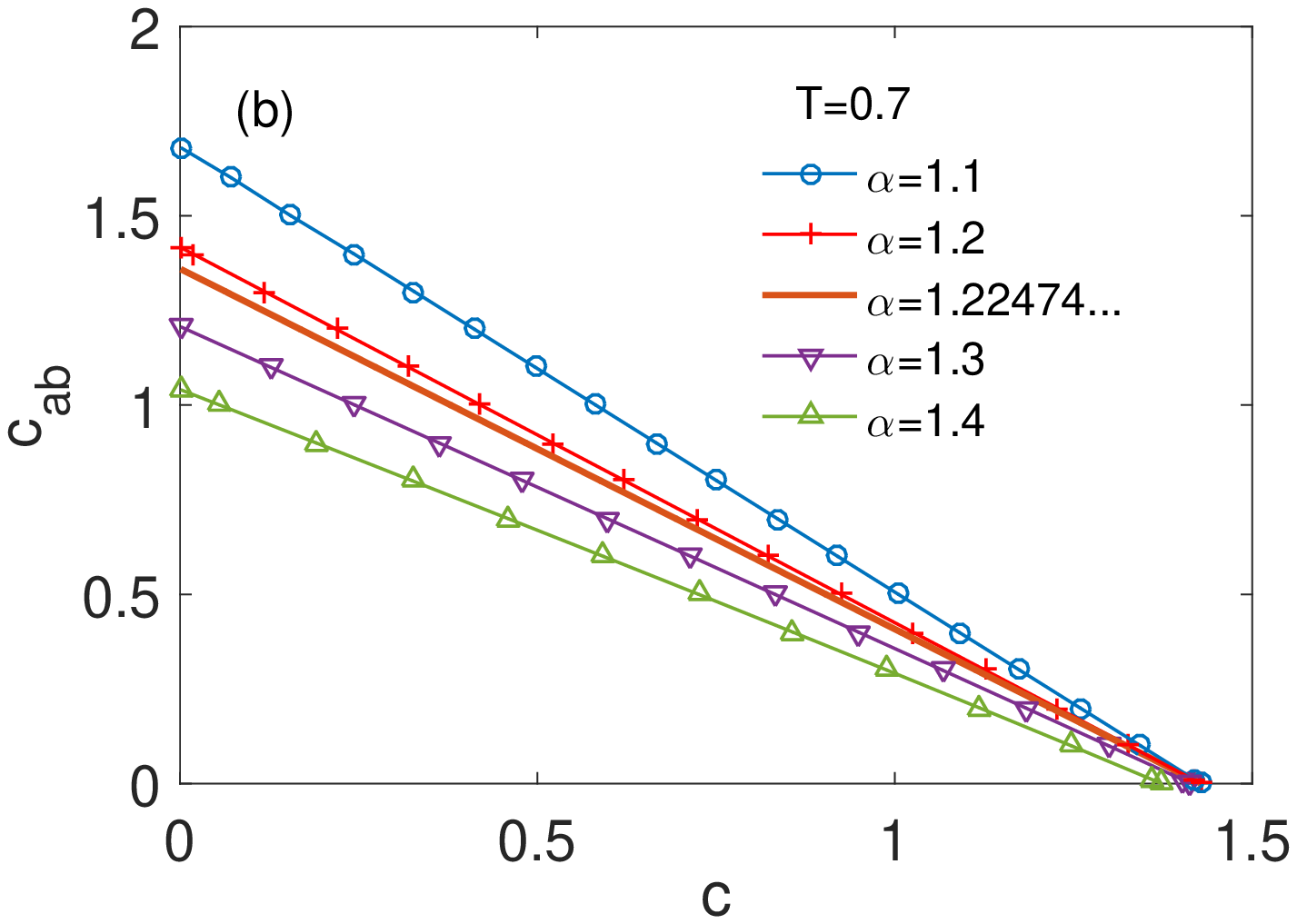}}
\end{center}
\caption{(color online). Lines of transition points for $(a)$ $T=0.4$ and $(b)$ $T=0.7$, on the plane $(c,c_{ab})$. The lines corresponding to $\alpha=1.1$ and $1.2$ indicate continuous transition points and lines with $\alpha=1.3$ and $1.4$, show discontinuous transition points. The solid line with $\alpha=1.22474...$ shows the line of tricritical points.
}
\label{f3}
\end{figure*}

\subsection{Erd\H{o}s--R\'enyi networks}
\begin{figure}[t]
\begin{center}
\scalebox{0.5}{\includegraphics[angle=0]{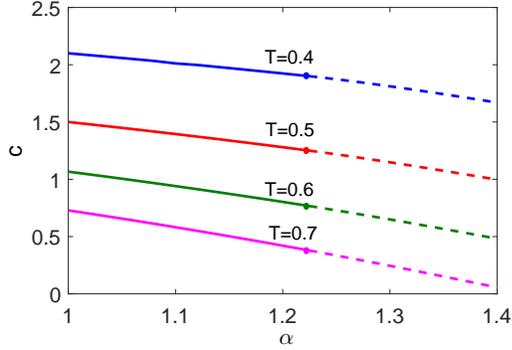}}
\end{center}
\caption{(color online). Phase diagram of the model on ER multiplex network for $c_{ab}=1$. Dashed lines indicate the positions of discontinuous phase transition points while solid lines represent continuous phase transitions. The tricritical point occurs at $\alpha=1.22474...$, for every value of $T$.
}
\label{f4}
\end{figure}
Let us consider multiplex networks such that each layer is an ER network with $P(k_i)=c_i^{k_i}e^{-c_i k_{i}}/k_{i}!$, for $i=a,b$ and $P(k_{ab})=c_{ab}^{k_{ab}}e^{-c_{ab} k_{ab}}/k_{ab}!$, where $c_i=\langle k_i\rangle$ and $c_{ab}=\langle k_{ab}\rangle$. For ER networks the generating functions are as $G_0(x)=G_1(x)= e^{-c(1-x)}$. Substituting $G_1(x)$ into Eq.~(\ref{eq101}) we find
\begin{eqnarray}
R(m, T_{ab})=\frac{(mT_{ab}c_{ab})^{m-1}e^{-mT_{ab}c_{ab}}}{m!}
\label{eq6}
.
\!\!\!\!\!
\end{eqnarray}
Also for ER uncorrelated networks, the degree distribution of supernodes is given as
\begin{eqnarray}
P_m(\textbf{q})=\frac{e^{-mc_a} (mc_a)^{q_a}}{q_a!}\frac{e^{-mc_b} (mc_b)^{q_b}}{q_b!}.
\label{eq8}
\!\!\!\!\!
\end{eqnarray}

For the sake of simplicity, let us consider the symmetric case $T_a=T_b=T$ and $c_a=c_b=c$. Also we assume $T_{ab}=\alpha ^{2}T^2$. In this case, $x_a=x_b\equiv x$, obtained from Eq.~(\ref{eq5}) for ER networks as,
\begin{eqnarray}
x=R_{\infty}+&&\sum_{m=1}^{\infty}R(m, T_{ab})\times\nonumber\\
&&\Big[(1-e^{-m \alpha c T x})^2+p(1-e^{-m c T x})e^{-mcTx}\Big].\nonumber\\
\label{eq9}
\!\!\!\!\!
\end{eqnarray}

Using Eq.~(\ref{eq6}), $R_{\infty}$ is given as
\begin{eqnarray}
R_{\infty}&&=1-\sum_{m=1}^{\infty}R(m, T_{ab})\nonumber\\
&&=1-\frac{1}{c_{ab}T_{ab}}\sum_{m=1}^{\infty}\frac{(m)^{m-1}(c_{ab}T_{ab}e^{-c_{ab}T_{ab}})^m}{m!}\nonumber\\
&&=1+\frac{W(-c_{ab}T_{ab}e^{-c_{ab}})}{T_{ab} c_{ab}}.\nonumber\\
\label{eq7}
\!\!\!\!\!
\end{eqnarray}
in which, $W(x)$ is the Lambert function.
Equation~(\ref{eq9}) can be simply rewritten in terms of the Lambert function, which enables us to solve the equation analytically,
\begin{widetext}
\begin{eqnarray}
x=&& \Psi(x)\equiv \nonumber\\
&&1+\frac{1}{T_{ab} c_{ab}}
\Big[-W(-c_{ab}T_{ab}e^{-c_{ab}T_{ab}-2\alpha c T x})
+2W(-c_{ab}T_{ab}e^{-c_{ab}T_{ab}-\alpha c T x})\nonumber\\
&&-p W(-c_{ab}T_{ab}e^{-c_{ab}T_{ab}{-}c T x})+p W(-c_{ab}T_{ab}e^{-c_{ab}T_{ab}{-}2cT x})\Big]\nonumber\\
\label{eq10}
\!\!\!\!\!
\end{eqnarray}
\end{widetext}
Equation~(\ref{eq10}) is a self-consistent equation for $x$ with parameters $c, c_{ab}, T$ and $\alpha$. To obtain the phase diagram of model, let us define $f(x)\equiv x- \Psi(x)$. Demanding that $f(x)=f^{'}(x)=0$ for $x>0$, we find the position of hybrid transitions, while a continuous transition occurs when $f(0)=f^{'}(0)=0$ and $f^{''}(0)>0$.

The lines in Fig.~(\ref{f3}) show the position of transition points in the plane $(c, c_{ab})$, for each values of $\alpha$. For $\alpha< 1.22474\ldots$, we find a line of continuous transition points. As $\alpha$ increases, the type of transition changes and the lines indicate the positions of hybrid transitions.
\begin{figure}[t]
\begin{center}
\scalebox{0.5}{\includegraphics[angle=0]{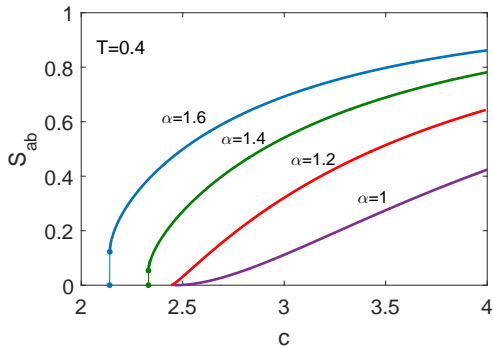}}
\end{center}
\caption{(color online). Relative size of the giant co-infected cluster vs mean degree $c$ for $c_{ab}=0.1$. Each curve corresponds to a different value of cooperativity.
}
\label{f5}
\end{figure}

Moreover, Fig.~(\ref{f4}) shows the phase diagram of the coinfection model, obtained for $c_{ab}=1$ and for different values of $T$. For low values of cooperation $\alpha$, the transition is continuous, while with increasing $\alpha$, the transition becomes discontinuous. The point $\alpha=1.22474\ldots$ is a tricritical point, determined by solution of $f(0)=f^{'}(0)=f^{''}(0)=0$.

Following the derivation of Eq.~(\ref{eq4}), we can obtain the probability that a given node belongs to the giant co-infected cluster:
\begin{eqnarray}
S_{ab}=1+\frac{1}{T_{ab} c_{ab}}\Big[&&-W(-c_{ab}T_{ab}e^{-c_{ab}T_{ab}{-}2\alpha c T x})\nonumber\\
&&+2W(-c_{ab}T_{ab}e^{-c_{ab}T_{ab}-\alpha c T x})\Big].\nonumber\\
\label{eq11}
\!\!\!\!\!
\end{eqnarray}
\begin{figure}[t]
\begin{center}
\scalebox{0.5}{\includegraphics[angle=0]{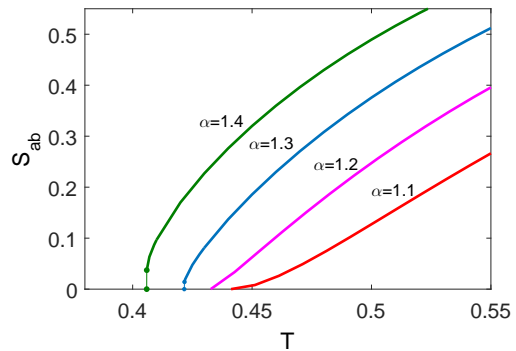}}
\end{center}
\caption{(color online). Relative size of the giant co-infected cluster for $c_{ab}=0.5$ and $c=2$ vs transmission probability $T$. Each curve corresponds to a different value of cooperativity.
}
\label{f77}
\end{figure}
We plotted $S_{ab}$ for different values of cooperation $\alpha$ in terms of the mean degree of networks, $c$ and the transmissibility, $T$ in Figs.~(\ref{f5}) and (\ref{f77}), respectively. For $\alpha> 1.22474\ldots$, the giant co-infected cluster emerges discontinuously at the transition point. As the fraction of overlapped edges is decreased, the jump values become smaller. However even for $c_{ab}=0$, the jump values are not zero and the transitions occur discontinuously. As the cooperativity increases, the epidemic threshold decreases to the smaller values of $c$ and $T$, which means that the cooperation between two diseases decreases the network's robustness against the propagation of both diseases.
\begin{figure}[t]
\begin{center}
\scalebox{0.5}{\includegraphics[angle=0]{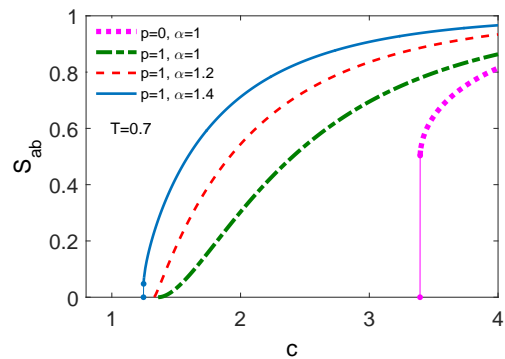}}
\end{center}
\caption{(color online). Relative size of the giant co-infected clusters with $(p=1, \alpha>1, T<1)$, compared with the size of the giant viable cluster $(p=0, \alpha=1)$, in ER multiplex network for $c_{ab}=0.1$. The curve corresponding to $(p=1, \alpha=1)$, shows the size of the giant infected cluster without cooperativity.
}
\label{f7}
\end{figure}

At the end, we compare the size of the giant co-infected cluster with the giant viable cluster. In multiplex networks, a viable cluster by definition is a set of nodes in which, for every type of edges, each two nodes are interconnected by at least one path following only (occupied) edges of this type. By this definition, the paths must pass through only the nodes in the cluster. Setting $p=0$ and $\alpha=1$ in Eqs.~(\ref{x_i}) and (\ref{eq4}), we can derive the size of the giant viable clusters. It was shown that viable clusters emerge discontinuously for every value of overlapping \cite{Havlin}. Figure~\ref{f7} shows the size of the viable cluster, compared with the co-infected clusters. The particular case of $(p=1, \alpha=1)$, is the overlap area of the giant connected components of two layers with occupied edges of types $a$ and $b$. The overlapped cluster include the viable cluster, since between each two nodes of the cluster, there is at least one path following each type of occupied edges, but the paths can pass through nodes outside of the cluster and then return to the cluster. The overlapped cluster, which corresponds to an infected cluster without cooperativity, emerges continuously. For $\alpha>1$, cooperation between two diseases occurs and diseases can infect a greater fraction of the nodes. With increasing values of $\alpha$, the size of the co-infected cluster becomes greater. Also, with increasing $\alpha$, the common terms in the derivation of the co-infected and viable clusters, make a higher contribution. Hence co-infected clusters show hybrid transitions similar to those seen for the viable clusters.

\section{Conclusion}
In this paper, we have introduced a coinfection model for
two diseases spreading on multiplex networks with the edges overlapping. Two diseases can propagate simultaneously on one multiplex network, such that both diseases can infect the same set of nodes. Our model illustrates how the existence of one infectious disease can enhance the propagation
of the other disease. Using the generating function technique, we have given an analytic solution for the size of the giant co-infected cluster, i.e., the set of nodes infected with both diseases, for uncorrelated multiplexes with arbitrary degree distribution. We have shown that the cooperation of two diseases decreases the network's robustness against propagation of both diseases, such that the epidemic threshold is shifted to smaller values of the edge transmission probability or the mean degree of networks. Our results show that for low cooperativity, the co-infected cluster emerges continuously. However, increasing the strength of cooperation, the type of phase transition changes to hybrid. Hence a tricritical point appears in our coinfection model.
We have compared the size of the giant co-infected cluster with the viable cluster for multiplex networks, considering edge overlapping. With increasing cooperativity, the co-infected cluster shows behavior similar to that of the viable cluster at the emergence point. The viable cluster is a subgraph of the co-infected cluster. However for large infected clusters these two clusters can coincide.

\begin{acknowledgments}
I would like to thank P. Grassberger for useful conversations and his suggestions for improving the manuscript.
\end{acknowledgments}


\begin{thebibliography}{99}

\bibitem{TB}
C. Kwan, J. Ernst, Clin Microbiol Rev 24(2):351 376 (2011).

\bibitem{HIV-HCV}
K. Lacombe et al., J. Int. AIDS Soc. {\bf 18} (Suppl. 4), 20479 (2015).


\bibitem{Rockstroh} J.K. Rockstroh and S. Bhagani, BMC Med. {\bf 11}, 234 (2013).

\bibitem{Brundage}
J.F. Brundage and G.D. Shanks, Emerg. Infect. Dis. {\bf 14}, 1193 (2008).






\bibitem{immune}
D. A. Vascoa, H. J. Wearinga, P. Rohani, Journal of Theoretical Biology {\bf 245}, 9-25 (2007).

\bibitem{Interaction}
M. E. J. Newman and C. R. Ferrario, PLoS ONE {\bf 8}, e71321 (2013).

\bibitem{Sanz}
J. Sanz, C.-Y. Xia, S. Meloni, Y. Moreno, Phys. Rev. X {\bf 4}, 041005 (2014).


\bibitem{Fakhteh}
L. Chen, F. Ghanbarnejad, W Cai and P. Grassberger, EPL, {\bf 104} 50001 (2013).

\bibitem{Cai}
W. Cai,	L. Chen, F. Ghanbarnejad and P. Grassberger, Nature Phys. {\bf 2015}, 3457 (2015).

\bibitem{clustercoin}
L. H\'ebert-Dufresne and B.M. Althouse, Proc. Natl. Acad. Sci. U.S.A. {\bf 112}, 10551 (2015).
%
\bibitem{Hanna}
H. Susi, B. Barre\'s, P. F. Vale, A.-L. Laine, Nature Commun. {\bf 6} 5975 (2015).
%
\bibitem{timescale}
M. Marv\'a, E. Venturino, R. B. de la Parra, J. of Appl. Math. {\bf 2015}, 275485, (2015).

\bibitem{SIRbond3}
P. Grassberger,  Math. Biosci. {\bf 63}, 157–172 (1982).

\bibitem{Buldyrev}
S. V. Buldyrev, R. Parshani, G. Paul, H. E. Stanley and S. Havlin, Nature {\bf 464}, 1025 (2010).

\bibitem{Son}
S.-W. Son, G. Bizhani, C. Christensen, P. Grassberger and M. Paczuski, Europhys. Lett. {\bf 97}, 16006 (2012).

\bibitem{avalanche}
G. J. Baxter, S. N. Dorogovtsev, A. V. Goltsev, and J. F. F.
Mendes, Phys. Rev. Lett. {\bf 109}, 248701 (2012).

\bibitem{Min}
B. Min and K.-I. Goh, Phys. Rev. E {\bf 89}, 040802(R) (2014).

\bibitem{multidir}
N. Azimi-Tafreshi, S. N. Dorogovtsev, and J. F. F. Mendes, Phys. Rev. E {\bf 90}, 90, 052809 (2014).

\bibitem{Grassberger}
P. Grassberger,  Phys. Rev. E {\bf 91}, 062806 (2015).


\bibitem{multilayer}
M. Kivela, A. Arenas, M. Barthelemy, J. P. Gleeson,
Y. Moreno, M. A. Porter, J.  Complex  Networks {\bf 2}, 203 (2014).


\bibitem{multiplex}
S. Boccaletti, G. Bianconi, R. Criado, C. I. del Genio, J. G\'omez-
Garde\~nes, M. Romance, I. Sendi\~na-Nadal, Z. Wang, and M.
Zanin, Phys. Rep. {\bf 544}, 1 (2014).

\bibitem{multiplex2}
K.-M. Lee, B. Min, and K.-I. Goh, Eur. Phys. J. B, {\bf 88} 48 (2015).

\bibitem{newman}
M. E. J. Newman, Phys. Rev. Lett. {\bf 95}, 108701 (2005).

\bibitem{Funk}
S. Funk and V. A. A. Jansen, Phys. Rev. E {\bf 81}, 036118 (2010).

\bibitem{Funk1}
S. Funk,  E. Gilad, and V. A. A. Jansen, J. Theor. Biol {\bf 264}, 501–509 (2010).

\bibitem{competition}
B. Karrer and M. E. J. Newman, Phys. Rev. E {\bf 84}, 036106 (2011).

\bibitem{idea}
Y. Wang, G. Xiao, J. Liu,  New J. Phys. {\bf 14} 013015 (2012).

\bibitem{modeling}
V. Marceau, P.-A. Noel, L. H\'ebert-Dufresne, A. Allard, L. J. Dub\'e, Phys. Rev. E. {\bf 84} 026105 (2011).

\bibitem{sahneh}
F. Sahneh, C. Scoglio, Phs. Rev. E {\bf 89}, 062817 (2014).

\bibitem{awareness}
C. Granell, S. G\'omez, and A. Arenas, Phys. Rev. Lett. {\bf 111}, 128701 (2013).

\bibitem{asymetry}
W. Wang, M. Tang, H. Yang, Y. Do, Y.-Ch Lai, and G. Lee, Nature Sci. Rep. Volume {\bf 4}, 5097 (2014).
%
\bibitem{memes}
X. Wei, N. C. Valler, B. A. Prakash, I. Neamtiu, M. Faloutsos,
and C. Faloutsos, IEEE J. Select. Areas Commun. {\bf 31}, 1049 (2013).

\bibitem{kermack}
Kermack W. O. and McKendrick A. G., Proc. R. Soc.
A, {\bf 115} 700 (1927).

\bibitem{SIRbond1}
M. E. J. Newman, Phys. Rev. E {\bf 66}, 016128 (2002).

\bibitem{Havlin}
Y. Hu, D. Zhou, R. Zhang, Z. Han, C. Rozenblat, and Sh. Havlin, Phys. Rev. E {\bf 88}, 052805 (2013).
%
\bibitem{linkoverlap}
B. Min, S. Lee, K-M. Lee, and K-I. Goh, Chaos Solitons Fractals {\bf 72}, 49–58 (2015).
%

\bibitem{componentsize}
M. E. J. Newman, Phys. Rev. E {\bf 76}, 045101(R) (2007).
%

\bibitem{GF}
M. E. J. Newman, S. H. Strogatz, and D. J.Watts, Phys. Rev. E
{\bf 64}, 026118 (2001).



\bibitem{SRG}
M. Z. Bazant, Physica A {\bf 316}  29–55 (2002).







\end{thebibliography}
\end{document}